\documentclass{article}
\usepackage{graphicx}
\usepackage{amsmath}

\input{tcilatex}
\input{tcilatex}
\begin{document}
\date{}
\title{{\flushright{\small IFUM-939-FT\\
            May 2009\\}}\vskip 0.5cm
Towards a more fundamental  theory  beyond 
quantum mechanics, avoiding  the Schroedinger paradox}
\author{{\small Francesco Caravaglios} \\
{\small Dipartimento di Fisica, Universit\`{a} di Milano, Via Celoria 16,
I-20133 Milano, Italy} \\
{\small and}\\
{\small INFN\ sezione di Milano}}
\maketitle

\begin{abstract}
The main distinction between classical mechanics and quantum mechanics is the 
lack in the latter of a full mechanical determinism: different final states 
can arise  from  the same physical state, after the measurement.
 No hidden variable is supposed to exist, nothing can discriminate 
 two apparently identical states even if they     give a different result. 

In this paper we try to put the basis for 
 a more fundamental theory that (approximately)
 coincides with quantum mechanics when comparing statistics, but it is more 
fundamental, since it  mathematically describes 
 measurement processes giving an explicit time evolution of the wave function 
during the collapse. The theory is deterministic even if the 
Heisenberg uncertainty principle  is still valid. The theory distinguishes 
physical states that collapse and physical states that do not collapse.
  The theory can be made compatible with all experiments 
 done in the past, but new phenomena such as violations of the Born law or the 
superposition principle could transpire.
However, even if we have  probably shown that it is possible to build 
{\it ad hoc} a theory that 
can describe both the wave function collapse and the Schroedinger linear
 evolution, a simple and unified construction is still missing.     
\end{abstract}

\section{Introduction}
Quantum Mechanics  (QM) is a probabilistic theory that predicts the time 
evolution
of any physical system, once some initial conditions have been assigned.
Even if these initial conditions appear to be unique\footnote{%
It is not possible to rule out the existence of hidden variables in the
initial conditions , that are physical objects which cannot be directly
fixed by any experimental instrument or measurement. But Quantum Mechanics
(QM) is a theory that does not consider this possibility.}, QM only provides
us with the probability that the physical system ends in one among a large
set of possible final states. We also know that the theory must correctly
describe experiments that show phenomena of destructive and constructive
interference. Equations are linear, and this linearity applies to any
physical system, both in first and second quantization\footnote{
For a short discussion of theories that violate the superposition principle 
see \cite{Caravaglios:2008ze}, 
where also potential risks, due to  such violations, are  
mentioned.}.
 This means that if a
physical system is described by the wave function $\psi (x,0)+\chi (x,0)$,
then the time evolution is $\psi (x,t)+\chi (x,t)$, where both $\psi (x,t)$
and $\chi (x,t)$ are solutions of the same Schroedinger equation. QM extends
this superposition principle to all systems, both microscopic systems like a
simple proton and macroscopic system like a gas of several atoms. As a
consequence, in real situations, QM predicts that a system quickly evolves
toward the linear superposition of extremely different physical states like
in the famous Schroedinger paradox.

Presumably, we can strictly apply QM to simple enough physical systems and
not any system. Probably the extent of validity of QM does not include
 physical objects containing several atoms.

In this paper we will try to put the basis for a mechanical and
deterministic theory that is approximately equivalent to QM in the
statistical sense, \textit{i.e.} it is almost compatible to QM. We require
mechanical determinism:\\ 
{\it 
the time evolution of any physical system  must satisfy the fundamental
principle\footnote{%
This principle is not true in QM, which predicts that the same state can
give different results after a measurement.} that different final states
always descend from  different initial states (initial conditions)}.
\\
 Therefore it is
necessary to assume that additional (hidden) variables distinguish two
states that evolve in different final states. In this work we restrict
ourselves to study those situations in first quantization, with the aim
to infer some necessary requirements to any algorithm that
simulates QM, both the Schroedinger linear time evolution and the wave
function collapse during the quantum measurement. To this purpose we will
make use of some stochastic methods.

\section{The Born rule of Quantum Mechanics and some possible violations}

The fundamental axioms of quantum mechanics are a direct consequence of the
following requirements

\begin{itemize}
\item[i)]  A probability distribution $P(x)$ is always positive.
\end{itemize}

This requirement can be easily realized, by choosing $P(x)=|\psi(x)|^2$,
where $\psi(x)$ is an arbitrary complex function. In general the function $%
\psi(x,t)$ depends on the time variable $t$, and assuming a linear time
evolution we have 
\begin{equation}
i\frac{\partial }{\partial \, t} \, \psi(x,t)=H\, \psi(x,t).  
\label{schroe1}
\end{equation}

 For any time $t$, we must have that

\begin{itemize}
\item[ii)]  The probability distribution always satisfies $\int P(x,t)\,dx=1$%
,
\end{itemize}

then $H$ must be Hermitian. The time evolution (\ref{schroe1}) is correct
before any quantum measurement occurs, that is the wave function collapse
into an eigenstate of an observable.

For the sake of clarity, we discuss the physical process of a measurement in
the specific case of a free particle confined by a wall barrier 
in a finite region $0<x<L$.
 To solve the differential equation (\ref{schroe1}), we
consider a lattice: we replace the real variable $x$, with an integer $0< n
\le L$ and take $L=6$, to make an analogy with a dice. This simplification
implies that the wave function $\psi(x,t)$ is replaced by a vector with six
complex components 
\begin{equation}
\vec{\psi}(t)=(\psi _{1}(t),\psi _{2}(t),\psi _{3}(t),\psi _{4}(t),\psi
_{5}(t),\psi _{6}(t)).   \label{vetto}
\end{equation}
The Schroedinger equation is 
\begin{equation}
i\,\ \frac{\partial }{\partial \,t}\,\ \psi (x,t)\,=\,-\frac{1}{2\,m}\,\frac{%
\partial ^{2}}{\partial \,x^{2}}\,\ \psi (x,t) 
\label{eqbuca1}
\end{equation}
that in our lattice approximation becomes 
\begin{equation}
i\,\ \frac{\partial }{\partial \,t}\,\ \vec{\psi}(t)\,\,=\frac{1}{2m}\left( 
\frac{6}{L}\right) ^{2}\left( 
\begin{tabular}{llllll}
2 & -1 & 0 & 0 & 0 & 0 \\ 
-1 & 2 & -1 & 0 & 0 & 0 \\ 
0 & -1 & 2 & -1 & 0 & 0 \\ 
0 & 0 & -1 & 2 & -1 & 0 \\ 
0 & 0 & 0 & -1 & 2 & -1 \\ 
0 & 0 & 0 & 0 & -1 & 2
\end{tabular}
\right) \,\ \ \ \vec{\psi}(t)\,   \label{eqbuca2}
\end{equation}
The equations (\ref{eqbuca1}) and (\ref{eqbuca2}) are equivalent in the
limit of lattice approximation that we are considering.

The equation (\ref{eqbuca2}) fixes the time evolution of the vector 
 $\vec{\psi}(t)$; but in what it follows, we replace the time derivative 
with a finite difference  $\vec{\psi}(t+ m\, \Delta)-\vec{\psi}(t+
 (m-1)\, \Delta)$: instead of  (\ref{eqbuca2}), a differential equation, we 
deal with an algorithm that gives $\vec{\psi}(t+ m\, \Delta)$ in terms 
of the preceding $\vec{\psi}(t+ (m-1)\, \Delta)$.
Any algorithm tends to a differential equation if the limit 
$\Delta \rightarrow 0$ exists. It is also true that any linear Schroedinger 
equation can be replaced by an algorithm that generates iteratively a sequence 
of  $\vec{\psi}(t+ m\, \Delta)$ for any integer $m$.\\

The vector $\vec{\psi}(t)$ contains six complex variables 
\begin{equation}
(\left| \psi _{1}\right| \,e^{i\,\alpha _{1}},\left| \psi _{2}\right|
e^{i\,\alpha _{2}},\left| \psi _{3}\right| e^{i\,\alpha _{3}},\left| \psi
_{4}\right| e^{i\,\alpha _{4}},\left| \psi _{5}\right| e^{i\,\alpha
_{5}},\left| \psi _{6}\right| e^{i\,\alpha _{6}})   \label{fasi}
\end{equation}
that satisfy the following condition 
\begin{equation*}
\sum_{i=1}^{6}\left| \psi _{i}\right| ^{2}=1.
\end{equation*}
However the vector (\ref{fasi}) is not enough to predetermine 
a single final state in  measuring processes. 
Additional information not contained in (\ref{fasi}) is
necessary if we require that different final states necessarily come from
different initial conditions.

Hereafter we give just one  example on how to make it possible, but it is 
understood that several and alternative mathematical representations
 can be used  to define the state of a physical system.

For example we can assume that a physical state 
is unambiguously defined once a sequence 
$\Gamma =\{x_{1},\cdots ,x_{N}\}$, of integer numbers $0<x_{n}\leq L$ ($L=6$
in our example) is fixed together with a set of (six) phases $\alpha _{i}$.
The sequence 
\begin{equation}
\Gamma =\{x_{1},\cdots ,x_{N}\}   \label{seq}
\end{equation}
plus the phases 
\begin{equation}
(\alpha_{1},...,\alpha_{L})
\end{equation}
 define only one\footnote{%
The converse is not true, since any permutation of the sequence $\Gamma 
$ gives the same vector (\ref{fasi}).} vector (\ref{fasi}) through the
identification 
\begin{equation}
p_i\equiv \left| \psi _{i}\right| ^{2}\equiv \frac{n_{i}}{N}   \label{ni}
\end{equation}
where $n_{i}$ is the number of times, the integer $i$ appears in the sequence 
$\Gamma $, and $N$ is the length of $\Gamma $. It is clear that our
definition of the physical state contains more (hidden) variables then the
vector (\ref{fasi}).

The time evolution both of the sequence $\Gamma(t)$ and of the phases $%
\alpha_i(t)$ unambiguously fixes the time evolution of $\vec{\psi}(t)$ 
(see (\ref{fasi})). 
If the probability distribution has no peak, and it is small
enough the time evolution of $\vec{\psi}(t)$ will probably follow a linear
Schroedinger evolution\footnote{%
But this is not mandatory, since in general hidden variables can play a not
negligible role in those situations too. We will not discuss this issue in
this paper.} 
\begin{equation}
i\,\ \frac{\partial }{\partial \,t}\,\ \psi (x,t)\,=\,\hat{H}\,\ \psi (x,t) 
  \label{eq20}
\end{equation}
but in some cases, probably when a measurement occurs, the evolution (\ref
{eq20}) fails, and one is forced to consider the full sequence $\Gamma(t)$
instead of just the vector $\vec{\psi}(t)$ to get the right and exact time
evolution. The aim of this work is to outline some crucial ingredients of the
 time evolution $\Gamma(t)$ that correctly describe the so called
Born rule of the quantum mechanics: a measurement of the observable $x$
induces a collapse of the wave function into an eigenstate of the observable 
$x$ with probability given by the squared wave function $|\psi(x)|^2$.

In our example this means that a measurement of $\vec{\psi}$
should induce a collapse into any of the following states (\thinspace $1\leq
i\leq 6$) 
\begin{equation}
(0,\dots ,e^{i\alpha _{i}},\dots ,0)
\end{equation}
with probability $p_i=|\psi _{i}|^{2}$. The following time evolution of $\Gamma
(t)$ will satisfy this Born rule.

\subsection{An algorithm that simulates the Born law of Quantum Mechanics}

For the sake of clarity we will make an analogy with a dice with six faces.
Exploiting this similarity we can find a rule that gives the sequence $%
\Gamma (t+m~\Delta )$ from the immediately preceding sequence, 
 that is $\Gamma (t+(m-1)~\Delta )$. Suppose that the six 
faces of the dice are not equally probable but at any time each face has
probability $p_{i}$, given by $p_{i}(t+(m-1)~\Delta )=n_{i}/N$. $n_{i}$ is
the number of times that $i$ appears in the sequence $\Gamma (t+(m-1)~\Delta
)$. Then we can get a new sequence $\Gamma (t+m~\Delta )$ simply throwing
this (non-equally probable) dice $N$ times.

We can repeat these steps, to obtain a new sequence $\Gamma (t+(m+1)~\Delta
) $, and taking into account that the dice face probabilities 
(see also eq.(\ref{ni}))
\begin{equation}
p_{i}(t+(m)~\Delta )=\frac{n_{i}}{N}\neq p_{i}(t+(m-1)~\Delta ) 
\label{itera}
\end{equation}
have changed since the sequence $\Gamma (t+(m)~\Delta )$ is changed.

In other words, throwing  the dice, we get 
a sequence $\Gamma $ and, in its turn, the
new sequence $\Gamma $ updates and changes the face probabilities\ $p_{i}\,\ $
of the dice \ (\ref{itera}). 
This iteration can be repeated several times:  
it can be shown that it exist a $M$ large enough
that for any $m>M$ we always get 
\begin{equation}
p_{i}(t+M\,\Delta )=\left\{ 
\begin{tabular}{ll}
1 & for $i=k$ \\ 
&  \\ 
0 & for $i\neq k$%
\end{tabular}
\right.   \label{deca}
\end{equation}
and $k$ is an integer between 1 and 6. It can be shown that the value of $k$
at the end of the process occurs a fraction of times proportional 
to the initial probability $p_{k}(t=0)$ .
Therefore this process simulates the Born rule but through a mechanical \
and deterministic process.
 In fact each  $\Gamma (t+m\,\Delta )$ derives from $\Gamma
(t+(m-1)\,\Delta )$ (simply throwing the dice).

\subsection{The Born rule through random-walk-like algorithms}

The simplest and more interesting method to simulate the Born rule in a
measurement process is obtained considering processes similar to a random
walk. Let us define what we mean for random walk process in this specific
context: the face probabilities $p_{i}(t+m\Delta )$ are obtained from those at
the preceding \ time $p_{i}(t+(m-1)\Delta )$ through a (pseudo-)random
algorithm, where for any $i$ 
\begin{equation}
p_{i}(t+m\,\Delta )=\left\{ 
\begin{tabular}{ll}
$p_{i}(t+(m-1)\,\Delta )+d_{i}\,\ g_{i}$ & if  $0<p_i<1$ \\ 
0 & if $ p_i<0 $\\ 
1 & if $ p_i>1$ \\
\end{tabular}
\right.   \label{deca1}
\end{equation}

where the $d_{i}<<1$ are some very small constants, while the $g_{i}$ are
stochastic variables that can take only two values +1 or -1, with equal
probability; they are only subjected to the following requirement 
\begin{equation*}
\sum_{i=1}^{6}\,d_{i}\,\ g_{i}=0.
\end{equation*}
It is possible to show that for any time evolution of this type, it exists a 
$M$ large enough for which the equation (\ref{deca}) applies and the final  
value $k$  is statistically distributed  as demanded by the
Born rule.

\subsection{When the wave function collapses into the eigenstate of an
observable}

In the previous section we reproduced the Born law, but we have not yet
clarified when a physical system is correctly described by the Schroedinger
equation and when the Schroedinger equation fails and it is replaced by a
more complex evolution that induces the wave function collapse. This
collapse involves the wave function of the full physical system, including
the experimental apparatus. We know that the Schroedinger equation is
correct during the time evolution when an atom does not interact with the
rest of the environment; on the other side we know that an electron that
travel across a bubble chamber leaves a track, and there is an elapsed time
during which the electron wave function irreversibly collapses and choose a
propagation direction. We talk about entanglement when the electron is not
an isolated system, since it interacts with several atoms. In the following
we will put the basis for a clearer mathematical distinction of the two
systems, in both scenarios described above.

\subsubsection{A metric in the configuration space to better define the
entanglement}

Hereafter  we would like to delineate  when the linear
superposition of two physically stable states does not give a new stable
state\ (as predicted by QM): instead this superposition immediately collapse
into one of \ them, due to a new dynamics that we usually call a measurement
process. The superposition principle is violated. In particular we would
like to know why states of  an electron propagating in
different directions can be superposed without inducing a collapse, while 
the superposition of two states representing a macroscopic object in two
completely different \ configurations inevitably collapses \ into one of
them (e.g. , alive or dead are very different configurations in the
Schroedinger paradox). \ Our goal  is to define a metric in the
configuration space, in order  to introduce a distance between any  couple
of physical states.

\subsubsection{The time evolution during a measurement: the entanglement and
possible violations of the Born law}

In this paragraph we will show a new algorithm in order to give a slightly
different  time law for the sequence $\Gamma (t+m\,\Delta )$ (defined in (%
\ref{seq})). The main difference is the physical variable $0<f<1$,  that
induces the wave function collapse if and only if $f$ is very close to 1. \
At each step of the iteration we cancel the first variable $x_{1}$ at the
beginning of  the sequence $\Gamma ,$ and we add a new variable $x_{\text{new%
}}$ \ at the end of $\Gamma ,$ as follows  
\begin{equation}
\Gamma (m\,\Delta )=(x_{1},\cdots ,x_{N})\,\ \ \ \Rightarrow \,\ \ \ \Gamma
((m+1)\,\Delta )=(x_{2},\cdots ,x_{N},x_{\text{new}}). 
\label{algo}
\end{equation}

The variable $1\leq x_{\text{new}}\leq 6$ is chosen at each step as follows 

\begin{eqnarray*}
x_{\text{new}} &=&x_{N}\text{ \ \ \ \ \ \ \ with probability  }f \\
&&\text{or} \\
x_{\text{new}} &=&i\text{ \ \ \ \ \ \ \ \ \ \ \ with probability  \ }%
p_{i}(m\Delta )\,\ \,(1-f)
\end{eqnarray*}
where we assume that $N$ is very large. $f$ introduces a correlation between 
$x_m$ and $x_{m-1}$, while this correlation is absent in the previous 
algorithm  described in section 2.1.  If $f>1-1/N$ then the sequence $%
\Gamma $ collapses into a sequence of all equal integers $i$ , after a large
 number of  iterations (\ref{algo}). The probability to get the
specific value $i$  at the end of the collapse is close to $p_{i}$, the
assigned  probability at the beginning of the process.

When $f\simeq 0,$  the probabilities $p_{i}$ \ are stable,  $p_{i}(m\Delta
)\simeq p_{i}((m-1)\Delta ),$ and the mechanism that induces a wave function
collapse is turned off. This means that the value of  $f$ is probably 
related to what has been  discussed in the previous subsection 2.3.1.

\section{Conclusion}

In this paper we have addressed the  well known  issue in quantum
mechanics, concerning the lack  of mechanical determinism: i.e. different
final states can derive from identical  initial quantum states. This lack of
determinism is ascribed to the measuring process, when a wave function
collapse occurs: one final state is selected among several  ones, with
apparently no a priori or theoretical  reason. However the Heisenberg 
uncertainty principle does not necessarily imply the absence of a more 
fundamental deterministic theory. The statistical intrinsic aspect of QM 
could  be due to our ignorance on some hidden variable dynamics.
  We have put the basis for a
theory that embeds quantum mechanics, but it is a more fundamental theory
since it  satisfies  mechanical determinism in all situations: different
final states always correspond to different initial conditions, probably 
due to   some hidden variables. 
 Quantum mechanics neglects this hidden variables
in the initial condition and only  deals  with the probability that a
certain final state occurs. 

First we have changed the definition of a physical state: instead of the
usual wave function $\psi (x),$ we have a sequence of real numbers $\Gamma
=\left\{ x_{1},...,x_{N}\right\} $ plus a phase $0<\alpha (x)<2\pi $ as a
function of $x.$ Once a  physical state is assigned according to the
previous definition, one (and only one)\ wave function can be deduced
through the following identification
\begin{equation}
\psi (x)\equiv \sqrt{n_{x}}\,\ e^{i\,\alpha (x)}   \label{wave}
\end{equation}
where $n_{x}$ is the probability density that  a randomly chosen  number,
extracted from  the sequence $\Gamma $ , is $x.$ The time evolution of the
physical state corresponds to the time evolution of the sequence $\Gamma (t)$
and the phase function $\alpha (x,t).$ In normal situations these evolutions
imply  the Schroedinger equation \ for the wave function (\ref{wave}), but
when a measurement occurs then the  wave function collapses and  the
Schroedinger equation fails. The exact dynamics now must take into account
the full sequence $\Gamma (t)$. 

We have addressed the issue to find  few explicit examples where the  time
law of $\Gamma (t)$ is such that the Born law of quantum mechanics finally
holds. We have not explicitly discussed this issue  in the paper, but we are
assuming that Einstein relativity is wrong and that a preferred reference frame
really exists. 
This  seems to be necessary, because we require time  causality,  a
mechanical and deterministic time evolution,  during which the wave function
collapses. Since the wave collapse is   {\it non local}, an absolute  (and
not relative) definition of time seems to be unavoidable\footnote{This is 
equivalent to assume the existence of  a preferred reference frame, 
where one  can define 
a universal absolute time. }.  However, even if we have  probably
 shown that it is possible to build 
{\it ad hoc} a theory that 
can describe both the wave function collapse and the Schroedinger linear
 evolution, a simple and unified construction is still missing.
    Experimental
searches should focus on the violations of the superposition principle
and/or the Born law. 

\end{document}